\documentclass{article}
\usepackage{arxiv}
\usepackage{url}
\usepackage{hyperref}
\usepackage{grffile}
\usepackage{graphicx}
\usepackage[numbers]{natbib}
\usepackage{amsmath}
\usepackage{amssymb}

\renewcommand{\headeright}{}

\newcommand{\beq}{\begin{equation}}
\newcommand{\eeq}[1]{\label{#1}\end{equation}}

\title{Improving data-driven global weather prediction using deep convolutional neural networks on a cubed sphere}

\author{
	\href{https://orcid.org/0000-0002-4789-7594}{\includegraphics[scale=0.06]{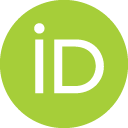}\hspace{1mm}Jonathan A. Weyn}\thanks{} \\
	Department of Atmospheric Sciences \\
	University of Washington \\
	Seattle, WA 98105 \\
	\texttt{jweyn@uw.edu} \\
	\And
	\href{https://orcid.org/0000-0002-6390-2584}{\includegraphics[scale=0.06]{orcid.png}\hspace{1mm}Dale R. Durran} \\
	Department of Atmospheric Sciences \\
	University of Washington \\
	Seattle, WA 98105 \\
	\texttt{drdee@uw.edu} \\
	\And
	Rich Caruana \\
	Microsoft Research \\
	Redmond, WA \\
	\texttt{rcaruana@microsoft.com}
}

\begin{document}
\maketitle

%
%

\begin{abstract}
We present a significantly-improved data-driven global weather forecasting framework using a deep convolutional neural network (CNN) to forecast several basic atmospheric variables on a global grid. 
New developments in this framework include an offline volume-conservative mapping to a cubed-sphere grid, improvements to the CNN architecture, and the minimization of the loss function over multiple steps in a prediction sequence. The cubed-sphere remapping minimizes the distortion on the cube faces on which convolution operations are performed and provides natural boundary conditions for padding in the CNN. 
Our improved model produces weather forecasts that are indefinitely stable and produce realistic weather patterns at lead times of several weeks and longer. 
For short- to medium-range forecasting, our model significantly outperforms persistence, climatology, and a coarse-resolution dynamical numerical weather prediction (NWP) model. 
Unsurprisingly, our forecasts are worse than those from a high-resolution state-of-the-art operational NWP system.
Our data-driven model is able to learn to forecast complex surface temperature patterns from few input atmospheric state variables. 
On annual time scales, our model produces a realistic seasonal cycle driven solely by the prescribed variation in top-of-atmosphere solar forcing. 
Although it is currently less accurate than operational weather forecasting models, our data-driven CNN executes much faster than those models, suggesting that machine learning could prove to be a valuable tool for large-ensemble forecasting. 

\end{abstract}

%
%

\section{Introduction}

Though still in its infancy, the application of machine learning (ML) to various aspects of weather forecasting is receiving increasing attention and yielding promising results.  Machine learning has been used in concert with output from numerical weather prediction (NWP) models in attempts to improve forecasts: \cite{Rasp2018a} used neural networks (NNs) to successfully improve post-processing of forecasts from general circulation models (GCMs) for surface stations; \cite{Rodrigues2018} demonstrated the ability of deep NNs to down-scale GCM output to higher horizontal resolution, and \cite{Scher2018} used NNs to estimate the uncertainty in weather forecasts.  Deep NNs have also been used to identify extreme weather and climate patterns in observed and modeled atmospheric states \citep{Liu2016,Kurth2019,Lagerquist2019}, to predict extreme weather events \citep[e.g.,][]{Herman2018}, and to provide operational guidance and risk assessment for severe weather \citep{McGovern2017}. \cite{Larraondo2019} developed deep NNs to extract spatial patterns in precipitation from gridded atmospheric fields, while \cite{Chattopadhyay2019} showed that deep NNs can skillfully predict extreme heat patterns several days ahead with relatively minimal input information.
Another machine-learning effort has focused on the improvement of physics parameterizations in GCMs for both weather forecasting and climate prediction \citep{Brenowitz2018,Rasp2018}.

With several decades of reliable weather data from satellite observations, widely available open-source software for machine learning, and efficient graphics processing unit (GPU) computing, recent studies have also begun to address the question of whether it is possible to forecast the weather using advanced ML algorithms such as deep learning to produce purely data-driven models, without explicitly enforcing the known physical laws governing atmospheric dynamics and physics. 
\cite{Dueben2018} used deep NNs trained on several years of reanalysis data to predict 500~hPa geopotential height on the globe at relatively coarse 6-degree resolution, demonstrating the ability of ML to produce modestly skillful atmospheric forecasts.
Using convolutional neural networks (CNNs), \cite{Scher2018a} and \cite{Scher2019} trained an algorithm on simulations from a simplified GCM that significantly outperformed baseline metrics and effectively captured the simplified-GCM dynamics. 

\cite{Weyn2019}, hereafter WDC19, trained CNNs similar to those of \cite{Scher2018a} and \cite{Scher2019} with over 20~years of historical reanalysis data to produce forecasts of 500~hPa height and 300--700~hPa thickness over the northern hemisphere. 
Their best CNN formulation was able to outperform a climatological benchmark for root-mean-squared error (RMSE) in the 500~hPa height field out to about 5~days of forecast lead time. 
However, the WDC19 model was applied only to the northern hemisphere on a latitude-longitude grid, and did not have appropriate boundary conditions at the north pole and the equator. 
In this study, we significantly improve on several aspects of the the previous best WDC19 model. Most notably, we use a volume-conservative mapping to project global data from latitude-longitude grids onto a cubed sphere, and design CNNs which operate on the cube faces, improving upon similar techniques used for processing $360^{\circ}$ imagery in the ML community \citep[e.g.,][]{Li2019}. 
The cubed-sphere mapping helps minimize distortion for planar convolution algorithms while also providing closed boundary conditions for the edges of the cube faces.
We further improve upon the CNN encoder-decoder architecture used in WDC19 and employ sequence prediction techniques to improve forecasts at longer time scales. Finally, surface-based atmospheric fields have been added to provide 
forecasts of surface temperature, a parameter of great importance in operational forecasting.

The remainder of this paper is organized as follows. 
In Section~2 we detail our new CNN-based weather forecasting model. 
The data and data processing are described in Section~3.
Results and evaluation of the model are presented in Section~4.
Finally, conclusions and discussion are provided in Section~5.

\section{The DLWP model}

As in WDC19, which introduced our Deep Learning Weather Prediction (DLWP) model, the model presented herein uses deep convolutional neural networks (CNNs) for globally-gridded weather prediction. 
A global weather prediction model must be given an initial multi-dimensional atmospheric state $\mathbf{x}(t)$ and yield the state of the atmosphere at a future time, $\mathbf{x}(t + \Delta t)$. 
To step the model forward in time, the predicted state must include all of the features of the input state. 
Dynamical models of the atmosphere compute tendencies of physical variables determined by equations of motion and physical parameterizations and then integrate forward in time. 
Following the methodology of WDC19, DLWP directly maps $\mathbf{x}(t)$ to an estimate of its future state $\mathbf{y}(t + \Delta t)$ by learning from historical observations of the weather. 
By feeding the predicted atmospheric state back as inputs to the model, DLWP algorithms can be iteratively propagated forward without explicitly using a numerical time-stepping scheme. 
As detailed in section~\ref{seqsection}, DLWP uses $\Delta t = 6$~h, a much larger time step than allowed for numerical stability in typical GCMs. 
In practice, and as detailed in WDC19, we find that model performance improves when using two consecutive atmospheric states $\mathbf{x}(t - \Delta t)$ and $\mathbf{x}(t)$ to predict two future states, $\mathbf{y}(t + \Delta t)$ and $\mathbf{y}(t + 2\Delta t)$.

This new work presents multiple significant improvements to the core DLWP model framework, which are detailed in turn in the following subsections. 
First, our model is adapted to operate on global data re-mapped to a cubed-sphere grid representation. 
Second, we use an improved neural network architecture based on the U-Net. 
Finally, we use sequence prediction techniques to improve DLWP for forecasts on medium-range and longer time scales.

\subsection{The cubed sphere in DWLP}

\subsubsection{Description of the grid}

One of the most natural coordinate systems for indexing data on the spherical Earth is  a latitude-longitude grid, but this system has singularities at the north and south poles that makes it difficult to use CNNs on this grid. A truncated expansion in spherical harmonic functions \citep{Durran2010} provides one elegant way to eliminate the polar singularities when approximating data on the sphere, but this representation, while potentially useful for deep learning on spherical data \citep{Cohen2018}, is inherently non-local and therefore not intuitive for applying CNNs to the gridded atmosphere. To preserve spatial locality we approximate data on the globe using the equiangular gnomomic cubed sphere.  This projection has been shown to give more uniformly-sized grid cells than the alternative gnomomic equidistant projection and to also produce better solutions to finite-difference \citep{Ronchi1996} and discontinuous Galerkin approximations \citep{Nair2005} to partial differential equations on the sphere.  The cubed sphere is used for state-of-the-art numerical weather prediction such as in the
FV3 dynamical core of the National Oceanic and Atmospheric Administration's Global Forecast System model \citep{Harris2013}.

\begin{figure}
	\centering
		\includegraphics[width=\textwidth, keepaspectratio]{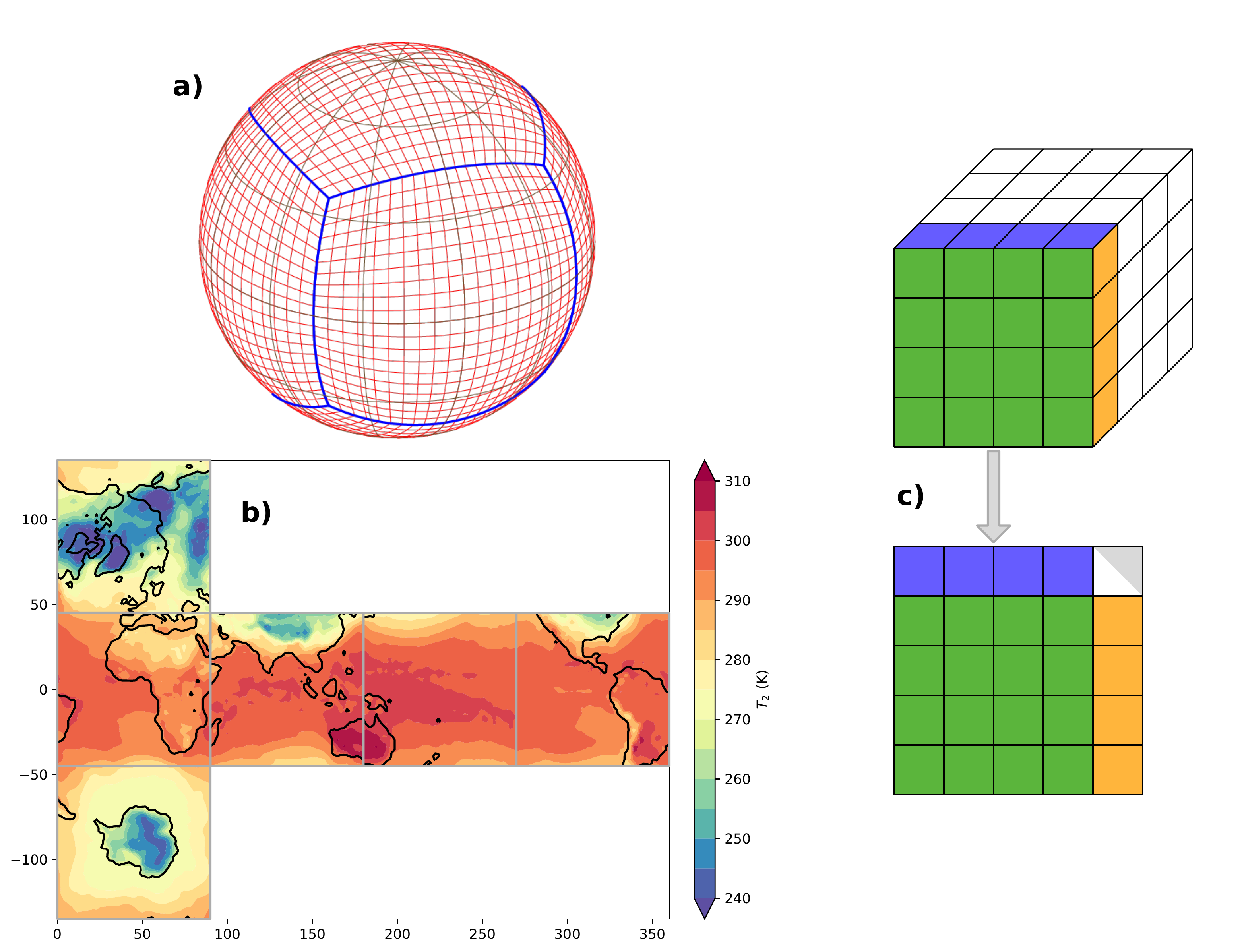}
		\caption{\label{cs} a) The gnonomic equiangular cube sphere grid with 20$\times$20 grid cells on each face, reproduced with permission from \cite{Purser2017}. Blue lines show the boundaries between faces; gray lines show latitudes and longitudes. b) Example 2-m temperature map for 00 UTC 5 Jan 2018 on the flattened cubed sphere. Gray lines outline individual faces of the cube. c) Points drawn for padding the green face with upper (blue) and right (orange) boundary conditions, after \cite{Eder2019}. The resulting flattened grid following the arrow shows that the corner point is ambiguous.}
\end{figure}

The gnonomic cubed-sphere geometry is illustrated in Fig.~\ref{cs}a. As an example, the field of air temperature 2 m above ground level is displayed on the six flattened cube faces in Fig.~\ref{cs}b.  Note that the construction of the cube faces ensures that each face has natural approximate boundary conditions provided by data in neighboring faces. 

In our DLWP application, all of the remapping between latitude-longitude coordinates and the the cubed-sphere grid is performed offline in the data pre-processing and post-processing pipeline. 
We use the Tempest-Remap library \citep{Ullrich2015,Ullrich2016} to perform both forward and inverse globally mass-conservative remapping. 
Transforming from the latitude-longitude grid to the cubed sphere yields a three-dimensional horizontal spatial grid where the first dimension indexes the six cube faces, and the other two dimensions provide an $x$-$y$-like indexing of the cells on each face. 

\subsubsection{The cubed sphere for convolutions in a CNN}

Convolution operations within the DLWP CNN are performed individually on each cube face, enabling us to use existing powerful software libraries for two-dimensional convolutions optimized for GPU hardware.
As an important consideration when applying cubed-sphere CNNs to weather prediction, DLWP learns separate weights and biases for the four faces centered on the equator and the two polar faces. 
Using one set of CNN weights for the equatorial faces and another for the poles enables the algorithm to reproduce the dramatically different evolution of weather patterns across the cube faces in those regions.
While the weights and biases for the Arctic face are identical to those for the Antarctic face, the construction of the cubed-sphere map shown in Fig.~\ref{cs}b results in atmospheric motions that are clockwise in the Antarctic and counterclockwise in the Arctic.
To reflect the change in the sense of cyclonic motion between the two poles, data on the Arctic face is flipped prior to applying each convolution operation, then flipped back.

It is necessary to pad the edges of the grid when performing a 2-D convolution operation with a filter size greater than $1\times1$ to avoid the loss of spatial dimensionality after the operation. 
We exploit this padding to create connections between the six faces by applying approximate boundary conditions from neighboring faces before each convolution operation to maintain continuity across the edges of the cube. 
The padding is illustrated in Fig.~\ref{cs}c. 
Padding points for the green (equatorial) face are drawn from the neighboring blue (polar) and orange (equatorial) faces. 
This process leaves the corner points ambiguous, a known issue for cube map convolutions \citep[e.g.,][]{Eder2019}. 
However, even in our complex weather prediction problem, this ambiguity does not appear to pose any problems for the CNN.  
On the equatorial faces, the ambiguous corners are drawn from the data on the polar faces (blue points) to maintain west-east periodicity. 
For efficiency, the corner points on the polar faces are filled using the same algorithm (fill upper and lower buffer rows first, then fill the left and right rows), however, the upper and lower sides are not meaningful distinctions on the polar faces, so the net effect of the corner-filling is more arbitrary. 
We did not find evidence that this efficient but arbitrary approach caused any difficulties.

\subsubsection{Applications of CNNs on the sphere in the literature}

Applying standard two-dimensional convolution operations within CNNs on faces of a cubed sphere is not unprecedented, as there are a number of examples of ``cube map'' convolutions applied to 360$^{\circ}$ images and videos within the computer science literature \citep{Ruder2017a,Monroy2018,Li2019}. 
Notably, \cite{Cheng2018} use a CNN applied to a cube map, along with padding the cube faces, to produce saliency maps from 360$^{\circ}$ images and videos, demonstrating that it is able to propagate saliency maps across edges of the cube map and outperform CNNs applied to standard equirectangular grids. 
\cite{Boomsma2017} apply a CNN with a cube map representation to classification of three-dimensional molecular models, likewise showing an improvement in performance over an equirectangular CNN. 
To the best of our knowledge, our DLWP on the cubed sphere represents the first application of CNNs on a cubed-sphere grid for a multi-dimensional regression problem and the first application of a volume-conservative cubed-sphere mapping for deep learning, and additionally offers the novel contribution of unique weights learned for the equatorial and polar faces.

There are also many other methods of sampling data on a sphere within CNNs that may have good performance in a data-driven weather prediction model. 
\cite{Cohen2018} developed convolution algorithms using spherical harmonics for rotation-invariant classification tasks. 
While elegantly suited for spherical data, regression of weather patterns requires fixed polar orientation of the globe and preservation of local interactions, which are not inherently accounted for in this method. 
Other studies have proposed several methods of applying convolutions on icosohedral grids, which promise to have less distortion than the cubed sphere \citep{Lee2018,Jiang2019,Zhang2019a}; however, these methods require complex adaptation of existing CNN software libraries and are likewise unproven for regression tasks. 
Finally, \cite{Coors2018} and \cite{Eder2019} propose alterations of convolution kernels that can be directly applied to equirectangular data but which account for the mapping between a latitude-longitude grid and the sphere. 
\cite{Coors2018} show that their method performs marginally better than a CNN on the cubed sphere.
While these techniques may be promising for future work, our choice of CNNs on a cubed sphere is motivated by physical constraints of the atmospheric system and by the successful application of cubed-sphere grids in operational NWP models; as will be discussed in section \ref{results}, it appears to perform very well.

\subsection{CNN architecture}

\begin{table}[t]
\caption{\label{unet-table} CNN architecture for DLWP as a sequence of operations on layers. The parameter $v$ represents the number of input fields, $t$ represents the number of input time steps, and $c$ represents the number of auxiliary prescribed inputs. The layer names (except for the suffix ``CubeSphere'') correspond to the names in the Keras library. The Concatenate layers append the states numbered in parentheses to the output of the previous layer.}
\begin{center}
\begin{tabular}{ccccc}
\hline
Layer & Filters & Filter size & Output shape$^a$ & Trainable params$^b$ \\
\hline
$input$ & & & (6, 48, 48, $vt + c$) & \\
\hline
Conv2D--CubeSphere & 32 & $3\times3$ & (6, 48, 48, 32) & 6,976 \\
Conv2D--CubeSphere (1) & 32 & $3\times3$ & (6, 48, 48, 32) & 18,496\\
AveragePooling2D & & $2\times2$ & (6, 24, 24, 32) & \\
Conv2D--CubeSphere & 64 & $3\times3$ & (6, 24, 24, 64) & 36,992 \\
Conv2D--CubeSphere (2) & 64 & $3\times3$ & (6, 24, 24, 64) & 73,856 \\
AveragePooling2D & & $2\times2$ & (6, 12, 12, 64) & \\
Conv2D--CubeSphere & 128 & $3\times3$ & (6, 12, 12, 128) & 147,712 \\
Conv2D--CubeSphere & 64 & $3\times3$ & (6, 12, 12, 64) & 147,584 \\
UpSampling2D & & $2\times2$ & (6, 24, 24, 64) & \\
Concatenate (2) & & & (6, 24, 24, 128) & \\
Conv2D--CubeSphere & 64 & $3\times3$ & (6, 24, 24, 64) & 147,584 \\
Conv2D--CubeSphere & 32 & $3\times3$ & (6, 24, 24, 32) & 36,928 \\
UpSampling2D & & $2\times2$ & (6, 48, 48, 32) & \\
Concatenate (1) & & & (6, 48, 48, 64) & \\
Conv2D--CubeSphere & 32 & $3\times3$ & (6, 48, 48, 32) & 36,928 \\
Conv2D--CubeSphere & 32 & $3\times3$ & (6, 48, 48, 32) & 18,496 \\
Conv2D--CubeSphere & $vt$ & $1\times1$ & (6, 48, 48, $vt$) & 528 \\
\hline
\multicolumn{5}{l}{\small $^a$Output shape is (face, $y$, $x$, channels).} \\
\multicolumn{5}{l}{\small $^b$Number of learned parameters for $t=2$, $v=4$, $c=4$. Total is 672,080.} \\
\end{tabular}
\end{center}
\end{table}

DLWP uses a fully-convolutional neural network to map the state of the atmosphere from one time step to the next. 
In simplistic terms, convolution operations within a CNN learn a prescribed number of $R\times S$ stencils, or ``filters,'' that are translated across an image (in this case a 2-D atmospheric field), producing an output image of each $R\times S$ input image area multiplied by the filters. 
The filter weights are learned by gradient back-propagation during training of the CNN.
Filters often extract certain types of features from the input images, such as edges or recognizable patterns. 
By also representing spatially-localized interactions -- that is, individual grid points in the output are only determined by the neighboring grid points within the convolutional stencil -- convolution operations are ideally-suited for recognizing spatial features in maps of the atmosphere and capturing localized advection. 
Nevertheless, large-scale processes are inherently accounted for by the sharing of convolutional filters across the input domain.
It is also worth noting that a fully-convolutional architecture has a relatively low number of trainable parameters: our DLWP CNN has about 700,000 parameters (see Table~\ref{unet-table}), while by comparison, a neural network which fully connects all input features to all output features would have 18 billion parameters, an untenable number. 

\begin{figure}
	\centering
		\includegraphics[width=\textwidth, keepaspectratio]{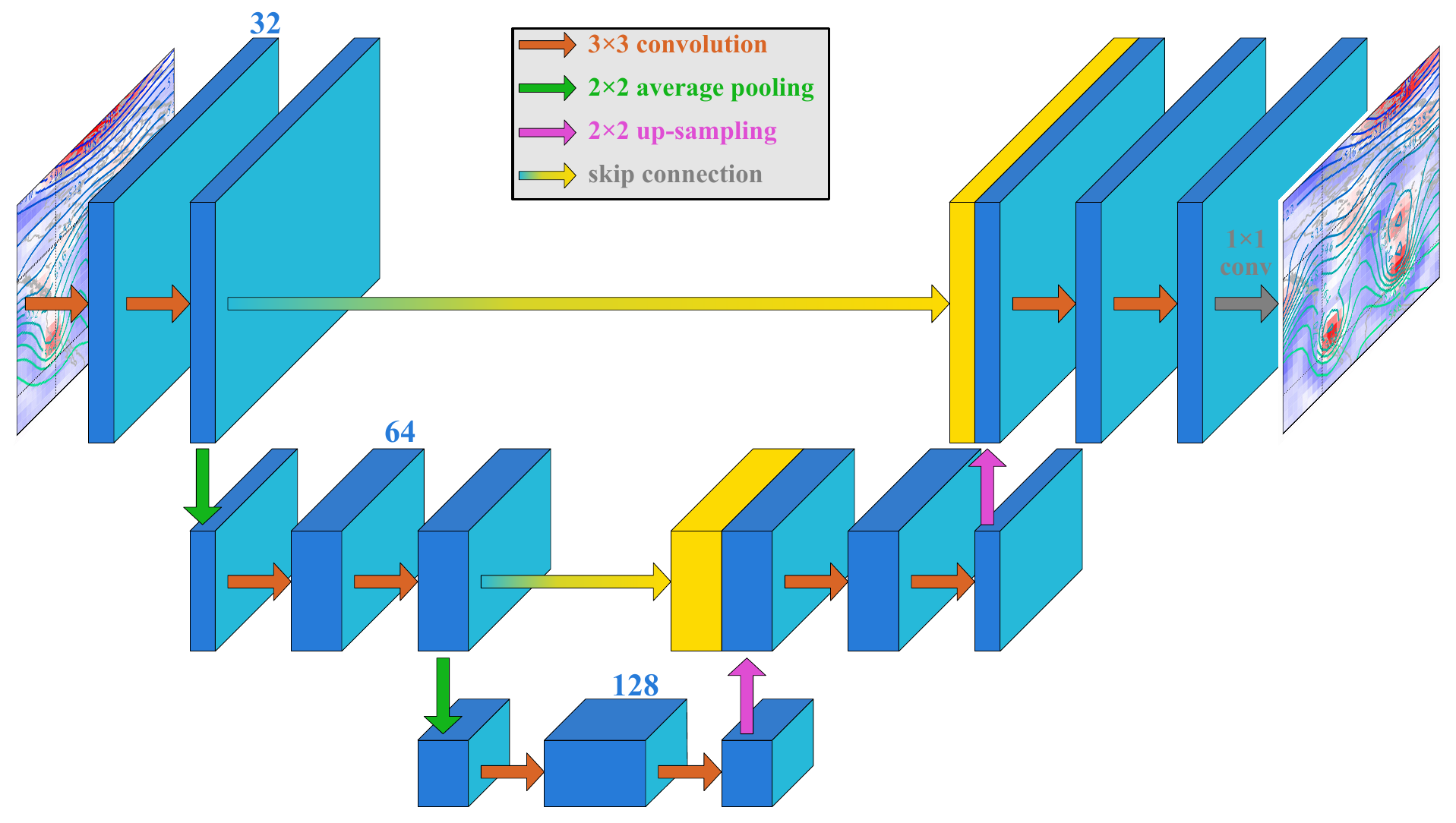}
		\caption{\label{unet} Schematic illustrating the architecture of our DLWP CNN based on the U-Net architecture. Each red arrow represents a 2-D convolution operating on each cube sphere face. Green and purple arrows indicate average-pooling and up-sampling operations, respectively. The blue-to-yellow lines represent skip connections, whereby the blue state is copied exactly to the yellow state vector and concatenated to the new blue state vector along the channels dimension. The final gray arrow is a $1\times1$ convolution. The blue numbers indicate the number of convolutional filters (channels) at each stage of the network (channel width is to scale).}
\end{figure}

The specific CNN architecture used herein is modeled on the popular U-Net architecture \citep{Ronneberger2015}, a variation on traditional encoder-decoder networks \citep{Baldi2012} that has shown good success in image segmentation tasks. 
\cite{Larraondo2019} tested several auto-encoder CNNs for the task of diagnosing precipitation from geopotential height fields in reanalysis data and found the U-Net to perform best. 
Figure~\ref{unet} shows our CNN architecture schematically. 
Each blue rectangle represents a state tensor at a stage of the CNN, as tabulated in Table~\ref{unet-table}.
In an encoder-decoder network, the first few convolutional operations (represented by red arrows in Fig.~\ref{unet}) in the network are followed by spatial pooling operations (green arrows) which reduce the spatial dimensionality of the state by a factor of 2 in both horizontal coordinates by taking an average value within each $2\times2$ sub-grid. 
By applying convolution operations with the same filter size ($3\times3$) on states with progressively coarser spatial resolution, the CNN is able to learn filters representing larger-scale atmospheric patterns.
The last few layers of the CNN are a mirrored up-sampling process (shown by purple arrows), whereby each spatial point of the image is copied to a $2\times2$ sub-grid, doubling the spatial dimensionality until the final convolutional operation yields an output state with the same dimensions as the input. 
This encoding-decoding process results in a loss of some spatial information, possibly resulting in a CNN prediction that is overly smoothed. 
To mitigate this, the U-Net concatenates the tensor state of the CNN at each encoding step to the tensor state immediately following each up-sampling operation in the decoding phase, thus allowing high-resolution information to flow freely through the network. 
The combination of multi-scale interactions from the encoder-decoder architecture and the skipped connections in the U-Net make this CNN architecture aptly suited for the complex, multi-scale weather prediction task.

Each convolution operation in the DLWP CNN, except for the final output layer, is followed by the application a nonlinear activation function $R(x)$, in our case a modified leaky rectified linear unit (ReLU), to each value $x$ in the state tensor such that
\begin{equation}
	R(x) = 
	\begin{cases} 
        0.1x & x \le 0 \\
	    x & 0 \leq x \leq 10 \\
		10 & x \geq 10,
	\end{cases}
\end{equation}
Capping $R(x)$ at a maximum value of 10 gave better results than the standard ReLU or leaky ReLU options, particularly for improving long-term forecast stability. 
The DLWP CNN is trained using the efficient Adam version of stochastic gradient descent optimization \citep{Kingma2014}, with a default learning rate of $10^{-3}$, and using mean-squared-error loss. 
To ensure that a suitable loss minimization is obtained, we train for a minimum of 100 epochs followed by early stopping conditioned on the validation set loss. 
If no new validation loss minimum is observed within 50 epochs, training stops and the model weights which yielded the smallest validation loss are restored. 

The DLWP CNNs are built using the open-source Keras library for Python \citep{keras} with Google's TensorFlow backend \citep{tensorflow}. 
We also acknowledge the open-source xarray project \citep{Hoyer2017} for much of the data processing pipeline.
The code for this project will be available upon publication at \url{www.github.com/jweyn/DLWP-CS}.

\subsection{\label{seqsection}Sequence prediction}

Based on existing NWP models it may seem intuitive to train a CNN to produce the best possible single-step forecast from a given atmospheric state, i.e., to minimize some error function $J\left[\mathbf{x}(t + \Delta t), \mathbf{y}(t + \Delta t)\right]$, where $\mathbf{y}(t+\Delta t)$ is the model prediction from the prior atmospheric state $\mathbf{x}(t)$, and $\mathbf{x}(t + \Delta t)$ is the correct evolved state from the training data. In 
practice this can yield a model that performs well for short-range forecasts but diverges from reality (or even blows up) for longer-range predictions \citep[e.g.,][]{McGibbon2019}. 
This is because there are no constraints on the CNN, physical or mathematical, that would prevent it from diverging from reality when its prediction fed back in as inputs no longer resembles an atmospheric state in the training data. 
In order to nudge the DLWP model towards learning to predict longer-term weather and improve its long-term stability, we train the model to minimize error on multiple iterated predictive steps using a multi-time-step loss function similar to the strategies in \cite{Brenowitz2018} and \cite{McGibbon2019}.
The loss function minimized over $T$ time steps may be written
\begin{equation}
	J_{\text{total}} = \sum_{n=1}^T \alpha_n J\left[\mathbf{x}(t + n\Delta t), \mathbf{y}(t + n\Delta t) \right],
\end{equation}
where $\alpha_n$ is an arbitrary prescribed weight. 
The predicted state $\mathbf{y}$ is an iterative sequence mapped by the DLWP CNN, denoted $M$, such that 
\begin{equation}
	\mathbf{y}(t + n\Delta t) = 
	\begin{cases} 
        M(\mathbf{x}(t)) & n = 0 \\
	    M(\mathbf{y}[t + (n-1) \Delta t]) & n \geq 1.
	\end{cases}
\end{equation}
The user-specified parameter $T$ can be tuned to find the best-performing model. 
Because the CNN training time scales with $T$ as a result of the iterative model mapping, we choose $T=2$ for computational efficiency. Our tests also gave some indication that using large values of $T$ can produce a model that makes overly smooth predictions tending towards climatology. 
The weights $\alpha$ are also adjustable should one choose to train a model that performs better on the earlier or later time steps, but for simplicity we let all $\alpha_n$ be 1. 
Finally, note that because one iteration of DLWP actually includes two 6-h time steps, the choice of $T=2$ implies that our model effectively minimizes the loss over a 24-hour forecast, equally weighting every 6-h snapshot of the evolving atmospheric state.

\section{Data}

The historical atmospheric data for DLWP is the European Centre for Medium-Range Weather Forecasts (ECMWF) ReAnalysis version 5 \citep[ERA5,][]{era5}. 
The data were retrieved through the Copernicus Climate Change Service (C3S) and re-gridded to a global 2-degree latitude-longitude grid through the Meteorological Archival and Retrieval System (MARS) toolkit. 
Data from 1979--2018 were retrieved every 3 hours. 
Note that while the ERA5 data were independently retrieved and processed, WeatherBench \citep{Rasp2020} also uses the same ERA5 data, so our results can be readily compared with those from other models using WeatherBench data.
Since the time resolution for the DLWP model is 6~h, there are samples in the data that only contain 00Z, 06Z, 12Z, and 18Z data and others which only have 03Z, 09Z, 15Z, and 21Z data. 
All data are re-gridded to a cubed sphere with 48 points on each side of the cube faces, which corresponds to roughly $1.9^{\circ}$ grid spacing in latitude and longitude in the center of the equatorial faces.

Data from 2017--2018 were set aside for the test set used in final model performance evaluation. 
We used the time periods from 1979--2012 for model training and 2013--2016 for model validation. 
Distinct periods for training, validation, and testing were selected to avoid including in the evaluation data times that have high correlation with neighboring times in the training data.
We assume that any climatological shifts in weather-pattern evolution over the 1979--2018 period are negligible. 

\subsection{Evolving variables and prescribed fields}

There are four two-dimensional input-output fields in the model: geopotential height at 500~hPa ($Z_{500}$), geopotential height at 1000~hPa ($Z_{1000}$), 300--700-hPa geopotential thickness ($\tau_{300-700}$), and 2-meter temperature ($T_2$). The geopotential height fields are vital to identifying the structure of mid-latitude weather systems, while $\tau_{300-700}$, which is dynamically related to mid-tropospheric temperatures, provides information about the growth and decay weather systems (WDC19). 
We include 2-m temperature as an impacts-based variable, which is vital for prediction of surface weather impacts such as heat waves, cold spells, and drought. Moreover, over the ocean the 2-m temperature is strongly influenced by the sea-surface temperatures (SST) and is therefore essentially a prediction of SST, which having longer time-scale variability, is important for sub-seasonal and seasonal forecasting.
Each variable is scaled by removing its global climatological mean and dividing by its global mean standard deviation. 
By scaling with the global mean we retain local spatial differences in variability, ensuring that the CNN loss function appropriately weights regions of high variability.
Since the CNN predicts scaled variables, an inverse scaling is applied to the model's output to forecast dimensional atmospheric variables. 

We also input three additional prescribed fields: top-of-atmosphere incoming solar radiation (insolation), a land-sea mask, and topographic height. 
These fields are not part of the model's output. 
Insolation is incorporated to give the model information about the diurnal and annual cycles, which are particularly important for predicting 2-m temperature. The land-sea mask is zero over ocean, one over land, and varies proportionately between 0 and 1 in coastal cells according to the fraction of their area covered by land. The topographic height is from ECMWF data regridded to the 2-degree latitude-longitude grid using the MARS toolkit.
In validation, adding these prescribed fields improved the performance of the model.

\subsection{Benchmarks\label{benchmarks}}

Benchmarks are necessary to contextualize the performance of DLWP. 
These benchmarks were inspired in part to facilitate comparisons with the aforementioned WeatherBench dataset \citep{Rasp2020}:
\begin{enumerate}
	\item climatology calculated relative to daily means from 1979--2010
	\item persistence
	\item a T42 spectral resolution version of the ECMWF Integrated Forecast System (IFS) model. This is a fully-dynamical atmospheric model with 62 vertical levels and an approximate horizontal resolution of $2.8^{\circ}$ in latitude and longitude near the equator and is thus slightly coarser than our DLWP model. The model was initialized with the same ERA5 data as our model, but on a coarser grid. Data were available for forecast lead times up to 7 days. The forecasts were initialized weekly within 2017--2018, with the first forecast at 00 UTC 1 Jan 2017, for a total of 105 forecasts. For every other week the initialization time is 12 UTC.
	\item a T63 spectral resolution version of the IFS model. Unlike the T42 IFS this version has 137 vertical levels and an approximate horizontal resolution of $1.9^{\circ}$ in latitude and longitude near the equator, thus being closer in horizontal resolution to our DLWP model. Also unlike the T42 IFS, this model was initialized with ECMWF analysis data and is coupled to an ocean wave model. As a result of the difference between initialization and verification data, this model has a noticeable error at early lead times. The T63 IFS was initialized at the same times as the T42 IFS, but forecast lead times up to 10 days were available.
	\item the operational subseasonal-to-seasonal (S2S) version of the ECMWF IFS. This is likewise a fully-dynamical model, with a fine horizontal resolution of 16~km that increases to 31-km after forecast lead times of 15 days. This model is fully-coupled to ocean and sea ice models, targeting predictions for time scales of 2 weeks to 2 months. This model is available as an 11-member ensemble but only the control forecast is evaluated. Like the T63 IFS model, some error at early lead times is is produced by minor differences between the initialization and verification data. The S2S model, 2018 version, is available twice weekly starting 00 UTC 1 Jan 2017 through 31 Dec 2017, and 1 Jan 2018 through 31 Dec 2018, for a total of 220 forecasts. All forecasts are initialized at 00 UTC, on dates of 1 Jan, 4 Jan, 8 Jan, and so on.
\end{enumerate}
The S2S model should produce better forecasts than our DLWP model because it makes predictions using many variables, many vertical levels, and high horizontal resolution. 
The T42 and T63 IFS models also include many variables and vertical levels, but are not at higher horizontal resolution than DLWP. 
An additional cautionary note applies for these IFS models: they use the same physics parameterizations for processes including radiation, convection, and boundary layer turbulence as the operational high-resolution IFS model run by ECMWF for medium-range forecasts. 
Such parameterizations are specifically tuned to perform well for the high-resolution operational model and therefore cannot be expected to perform well within the lower-resolution IFS simulations. 

To mitigate differences in model grids, all of the forecasts and the verifying ERA5 data were forward mapped from the latitude-longitude resolution at which they are provided to the cubed sphere and then inverse mapped back to the regular $2^{\circ}$ latitude-longitude grid. 
This allows a uniform comparison to our DLWP model which operates on the cubed-sphere grid. 
Our DLWP model is also initialized at the same times as the S2S operational model for direct comparison. 
Despite slightly different initialization times for the T42 and T63 IFS models, because we average at least 100 forecasts across all seasons for each model, the sample is representative of the average distribution of forecasts across the test set.

\section{Results\label{results}}

In the following section, we detail the spatially- and temporally-averaged error in DLWP and the benchmark forecasts as a function of forecast lead time.  We then examine the structure of several key forecast fields in an example 4-week forecast.  Finally, to evaluate the long-term behavior of our forecasts, we consider the evolution of the zonally-averaged 500 hPa geopotential from a free-running one-year forecast.

\subsection{Globally averaged forecast error}

\begin{figure}
	\centering
		\includegraphics[width=\textwidth, keepaspectratio]{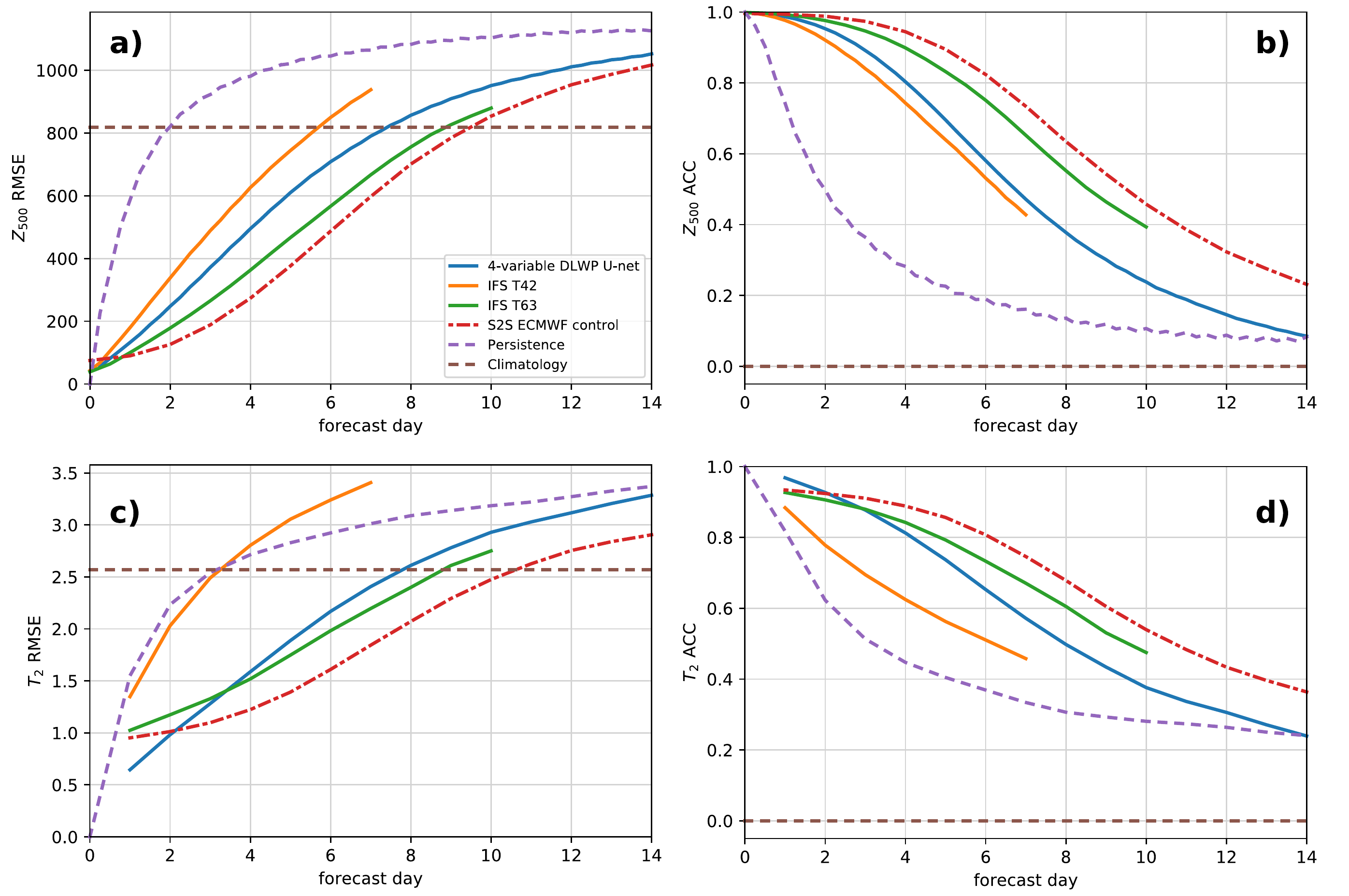}
		\caption{\label{error} Forecast error for DLWP (blue lines) and all of the benchmarks, as labeled, as a function of forecast lead time during 2017--2018. The error is globally-averaged and area-weighted in latitude. a) Root-mean-squared error in $Z_{500}$. b) Anomaly correlation coefficient in $Z_{500}$. c) Root-mean-squared error in daily-averaged $T_2$. d) Anomaly correlation coefficient in daily-averaged $T_2$.}
\end{figure}

The DLWP model and the benchmarks are evaluated using two key measures for forecast accuracy, both of which are used to assess the performance of state-of-the-art operational NWP models \citep{Vitart2004}. 
The root-mean-squared error (RMSE) of a forecast vector $\mathbf{f}(t)$ at some time $t$ is defined as
\begin{equation}
	\text{RMSE} = \sqrt{\overline{\left(\mathbf{f}(t) - \mathbf{o}(t)\right) ^ 2}},
\end{equation}
where $\mathbf{o}(t)$ is the observed state, the overbar denotes a spatial average, and the dot product of a vector with itself is denoted as the square of the vector. The RMSE gives a good point-by-point metric of the accuracy of a forecast. 
The other metric used for evaluation is the anomaly correlation coefficient (ACC), defined as
\begin{equation}
	\text{ACC} = \frac{\overline{\left(\mathbf{f}(t) - \mathbf{c}(t)\right)\cdot \left(\mathbf{o}(t) - \mathbf{c}(t)\right)}} {\sqrt{\overline{\left(\mathbf{f}(t) - \mathbf{c}(t)\right) ^ 2} \;\; \overline{\left(\mathbf{o}(t) - \mathbf{c}(t)\right) ^ 2}}},
\end{equation}
where $\mathbf{c}(t)$ is the climatological value for the verification time (daily climatological values are used herein). 
A perfect forecast has an ACC score of 1, while a score of 0 indicates a forecast with no skill relative to climatology. 
Where the RMSE penalizes large differences in actual forecast state, the ACC penalizes incorrect patterns of anomalies and is agnostic to the magnitude of the anomalies. 
The ACC is able to distinguish whether a forecast has meaningful spatial patterns instead of producing smooth states that resemble climatology.

Let us first examine the globally averaged forecast errors in the 500-hPa height field, which are plotted as a function of forecast lead time up to two weeks in Fig.~\ref{error}.
As noted in Section~\ref{benchmarks}, these errors are averaged over more than 100~forecast initialization times in the test set (2017--2018). 
The RMSE in $Z_{500}$ (Fig.~\ref{error}a) shows that DLWP significantly outperforms persistence at all lead times, climatology out to more than 7~days, and the T42 IFS at all available lead times. 
On the other hand, DLWP is outperformed by the T63 IFS and the operational S2S, the latter of which has errors that are lower than climatology out to more than 9~days. 
It is not particularly surprising that the S2S and even the T63 IFS perform well because $Z_{500}$ is not difficult to forecast with a state-of-the-art dynamical NWP model.
The best DLWP model in WDC19 beat climatology up to a lead time of about 5~days (WDC19, Fig.~6). Measured by the lead time up to which a model beats climatology,
our improved DLWP model (7-day lead time) has approximately halved the forecast skill deficit relative to state-of-the-art operational models (9-day lead time). 

The ACC scores for $Z_{500}$ (Fig~\ref{error}b) give rankings similar to those from the RMSE scores for the DLWP model relative to the benchmarks, with the exception of climatology which, by definition, has a score of zero.
The forecast horizon for an ACC score of 0.5 is more than 2~days longer for the operational S2S model than our DLWP model, with the value for DLWP dropping below the 0.5 threshold just shy of 7~days and the S2S reaching that mark at 9.5~days. 
DLWP does not have as much of an advantage over the T42 IFS in the ACC score, but still comfortably outperforms persistence forecasts. 
There is also slightly larger separation between the DLWP model and the T63 IFS in the ACC score, with the latter having the advantage. 
Nevertheless, the good ACC score for DLWP indicates that it is producing  spatial weather patterns with reasonable disturbance amplitudes rather than overly smooth forecasts approximating climatology.

Shifting our focus to the 2-m temperature metrics shown in Fig.~\ref{error}c,d, we again see a fairly similar performance ranking among the models. 
First, we note that, in order to match the available products from the operational S2S dataset, the errors are calculated for daily mean $T_2$. 
Additionally, errors at early lead times in the IFS and S2S models are to a some extent due to differences in the initialization data (T63 IFS, S2S) or model grids (T42 IFS), which are pronounced in a highly spatially-varying field such as $T_2$. 
As measured by the RMSE (Fig~\ref{error}c), DLWP clearly outperforms persistence and the coarse-resolution T42 IFS; the DLWP-model errors remain lower than those of climatology until the 8-day mark. 
As was the case for $Z_{500}$, the T63 IFS performs slightly better than the DLWP model, beating climatology up to the 9-day mark. The RMSE of the best model, the S2S remains lower than climatology until 11~days out. 
In terms of ACC scores (Fig~\ref{error}d), all of the models perform notably better than persistence and retain good forecast skill relative to climatology. 
The relative ranking of model skill again has the DLWP model clearly beating the T42 IFS, but performing worse than the T63 IFS and the operational S2S model.
The DLWP model exceeds the 0.5 skill threshold out to 8~days while the S2S model does so out to 11~days. 

\subsection{A typical forecast state at 4-week lead time}

\begin{figure}
	\centering
		\includegraphics[width=\textwidth, keepaspectratio]{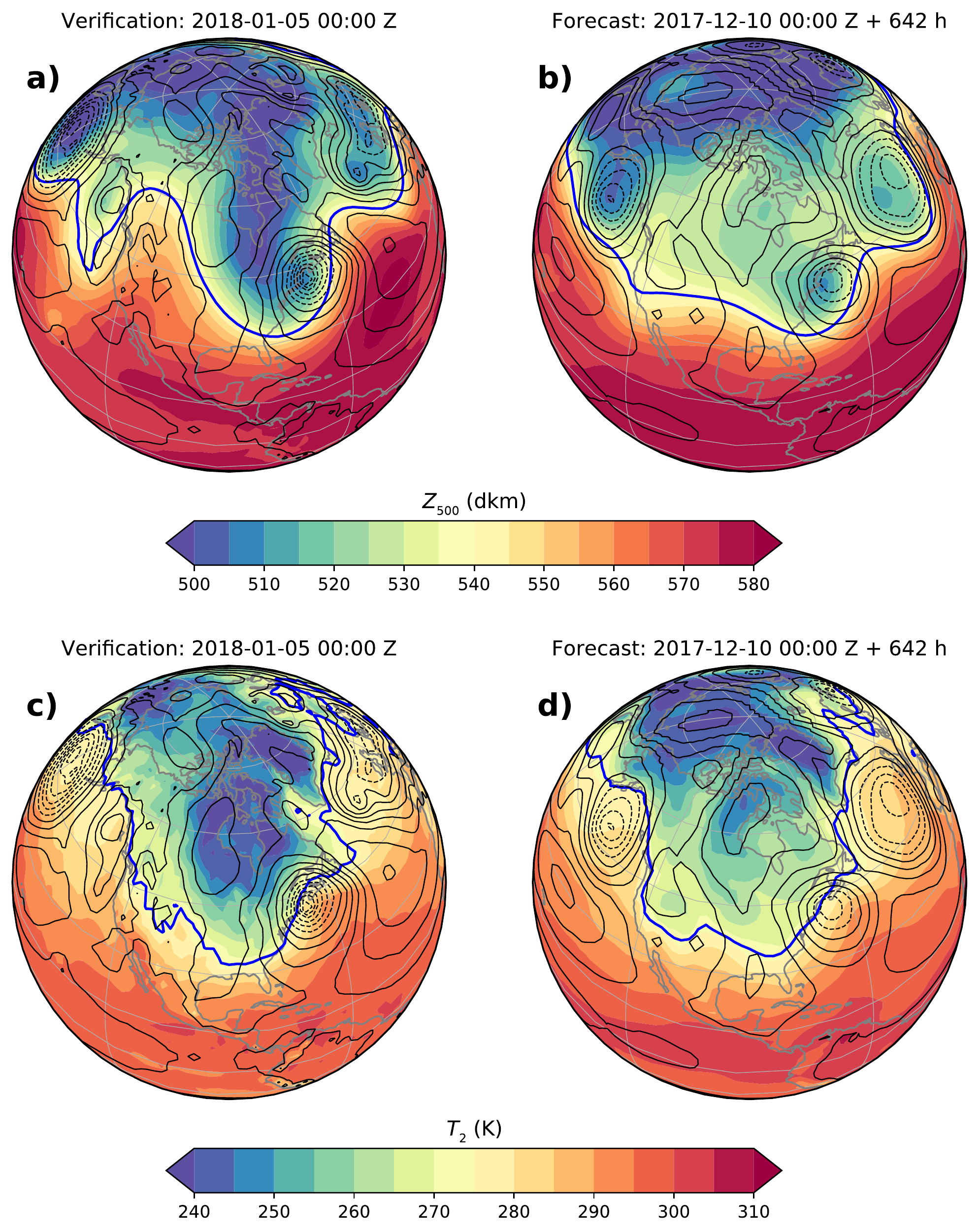}
		\caption{\label{fcst} Example observed atmospheric state for an active weather pattern with a strong cyclone off the eastern US (a,c) and a similar example DLWP forecast state (b,d). The forecast is 642~h out initialized 00 UTC 10 Dec 2017 and valid 18 UTC 5 Jan 2018, while the observation is for 00 UTC 5 Jan 2018. a,b) The color shading is $Z_{500}$, with $Z_{1000}$ in the black contours (contoured every 100~m with negative contours dashed). The 540-dam $Z_{500}$ line is shown in blue. c,d) $T_2$ is in the shaded color, with the $0^{\circ}$C isotherm in the blue line. Black contours are as in a,b).}
\end{figure}

Unlike our earlier  DLWP models which represent the northern hemisphere using cylindrical geometry (WDC19), the current DLWP model is a true global model.  It is therefore possible to generate arbitrarily long free-running forecasts from a single initial state without suffering from lateral-boundary-condition errors at the equator or the poles. 
Every one of the four-week forecasts initialized twice weekly in the two-year test set (210 total forecasts) was free from instabilities and the amplification of spurious perturbations.  
This is remarkable considering the difficulties that have been previously encountered in creating stable models of atmospheric flows from purely data-driven techniques \citep{Dueben2018} or with data-driven algorithms inserted as parameterizations in GCMs \citep[e.g.,][]{Brenowitz2018}. 

In this section we assess the realism of the forecast fields at multi-week lead times. We selected 00 UTC 5 Jan 2018 as an interesting wintertime reference state with high-amplitude perturbations over the North American region. As shown in by the $Z_{500}$ and surface pressure fields in Fig.~\ref{fcst}a, there is a deep trough and a strong surface cyclone off the east coast of the United States.  A cold-air outbreak is apparent in the  2-m temperature field beneath the trough (Fig.~\ref{fcst}c).  
We saw in the previous section that DLWP produced modestly skillful spatial patterns (good ACC scores) in forecasts up to two weeks. 
How well can DLWP reproduce high-amplitude patterns such as that in Fig.~\ref{fcst}a,c in forecasts at lead times of two weeks and longer?  

To answer this question, the full two years of forecasts from the test set were examined to find the single forecast at lead times greater than 2~weeks having the lowest RMS difference with respect to the reference-state $Z_{1000}$ field over the North American sector.  The closest match (minimum RMS difference) was the 642-h forecast initialized at 00 UTC 10 Dec 2017 and verifying at 18 UTC 5 Jan 2018, 18 hours after the time of the observed state. As might be expected from the selection criteria, features common to both the forecast and the reference state in the are apparent in the $Z_{1000}$ field, including the cyclone off the east coast of the US. The forecast cyclone is weaker than observed, but is still accompanied by a realistic warm sector in the 2-m temperature field (Fig.~
\ref{fcst}d).  Over the eastern US, warmer temperatures would be expected at 18 UTC than at the 00 UTC time of the reference-state.  Allowing for this difference, the $T_2$ fields compare reasonably well.  Deterministic forecasts have no skill at lead times of 26.75 days, so the quality of the match is serendipitous---indeed lead times shorter than this, but still longer than two weeks, did not produce better matches.

The amplitudes of the troughs and ridges in the forecast $Z_{500}$ field (Fig.~\ref{fcst}b) are considerably weaker than those in the reference state (Fig.~\ref{fcst}a).  A better match might have been found using RMS differences in $Z_{500}$ instead of $Z_{1000}$, but there is a tendency for the model to produce overly smooth fields, and we therefore believe the current example meets our goal of illustrating the nature of the atmospheric states predicted by the DLWP model at long lead times.  Our DLWP forecasts also tend to underestimate  the development of blocking patterns associated with strong ridges over the western US (not shown). 
Such blocking patterns are very important large-scale weather phenomena that can cause droughts and heat waves in the US, hence it is important for a model to be able to capture them. 
Despite the impressive performance of DLWP, there is clear room for improvement in longer-term forecasts, with particular emphasis on the ability of the model to produce extreme weather patterns.

\subsection{The zonally averaged evolution of a one-year forecast}

\begin{figure}
	\centering
		\includegraphics[width=\textwidth, keepaspectratio]{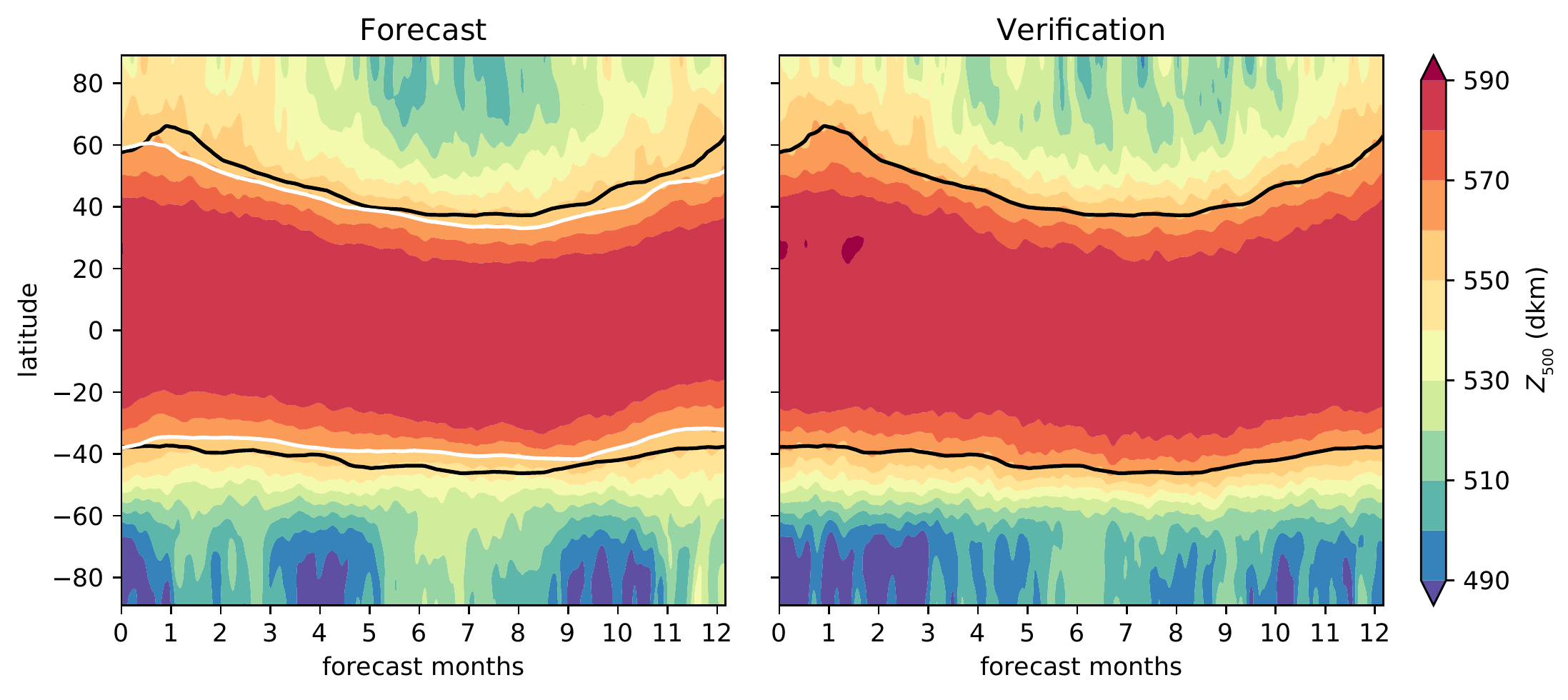}
		\caption{\label{climo} Zonal-mean $Z_{500}$ as a function of forecast lead time for a) a DLWP forecast initialized in July and b) the verifying observations. A running mean of 3 days has been applied to smooth the data in the colored contours. The black (white) lines are the 560-dam line from the verification (forecast) with a 15-day centered running mean smoothing.}
\end{figure}

In this section we  consider the behavior of a free-running one-year forecast initialized from data at the two time steps 18 UTC July 3, 2017 and 00 UTC July 4, 2017.
Figure~\ref{climo} shows the zonal-mean 500-hPa geopotential as a function of time and latitude from this 1-year DLWP simulation along with the corresponding zonal-mean field from the ERA5 reanalysis. 
Based only on seasonal variation in the top-of-atmosphere insolation and the immediately preceding atmospheric states, the DLWP model is clearly able to capture the basic structure of the annual cycle, with lower values of $Z_{500}$  near the north pole during the winter months followed by a subsequent increases during spring and the onset of the next summer, in approximate agreement with the true annual cycle.  

The true progression of the annual cycle in mid-latitude $Z_{500}$ is indicated by the 15-day running mean of the 560-dam geopotential height contour in the ERA5 dataset (black line) in both panels. The 15-day running mean of the 560-dam contour in the DLWP forecast is indicated by the white lines in Fig.~\ref{climo}a.  After the first two weeks, the location of the 560-dam contour is biased equatorward, although particularly in the northern hemisphere, the north-south seasonal displacement of 560-dam contour in the DWLP forecast does follow that of the ERA5 data reasonably well.  Also evident in Fig.~\ref{climo} is the weaker temporal variability in the weather in the DLWP model compared to the observations, as indicated by the reduced waviness of contour lines in the latitude band 20--60$^{\circ}$. 

Similar annual cycles in $Z_{500}$ were generated by additional forecasts initialized in different months of the year (not shown). 
Thus, while the annual cycle in these free-running DLWP forecasts is certainly not perfect, it is impressive that the model remains stable and produces an approximately correctly response to the seasonal changes in the top-of-atmosphere insolation.

\section{Discussion and Conclusions}

In this paper, we have extended our previous CNN-based DLWP model (WDC19), which predicted the evolution of northern-hemisphere 500-hPa geopotential height and 300--700-hPa thickness, to a full global forecast and added a pair of additional forecast fields (1000-hPa geopotential height and 2-meter temperature) along with three additional prescribed inputs (top of the atmosphere insolation, a land-sea mask and topographic height). 
We also made three important improvements to the model architecture: (1) global data are represented on the cubed sphere for which 2-D convolutions can be naturally computed on the cube faces, (2) an additional convolutional layer is employed before each average-pooling or up-sampling step along with U-net skip connections, and (3) a multi-time-step loss function is used to improve the stability and accuracy of long-term forecasts. 
Free-running one-year forecasts are stable and provide realistic characterizations of atmospheric states, albeit with modest differences from observed weather.

This new DLWP model clearly outperforms a coarse-resolution T42 configuration of the ECMWF IFS dynamical model, both with respect to global averaged RMS error and the anomaly correlation coefficient.  The T42 IFS model forecasts three-dimensional fields of horizontal velocity and temperature on 62 vertical levels along with the surface pressure.  Not counting six other fully 3D prognostic fields related to moisture, clouds and precipitation, or other 3D diagnostic fields like the vertical velocity, the T42 IFS forecast steps forward 187 spherical shells of data.  In contrast our DLWP model steps forward 8 spherical shells of data (four variables at two time levels) in 12-hour time steps. 
Despite our model having higher horizontal resolution (roughly $1.9\times1.9^{\circ}$ in latitude and longitude) than the T42 IFS (at $2.8\times2.8^{\circ}$), it is surprising that the full-physics model is inferior to our DLWP model forecasts of a relatively smooth field such as the 500-hPa height.  It is also interesting that the DLWP model considerably outperforms the T42 IFS forecasts of 2-m temperature.  On one hand this might be expected because 2-m temperature should be a difficult field for the T42 IFS to capture due to its coarse resolution and its use of physical parameterizations optimized for much finer grid cells.  But on the other hand, the DLWP model had to overcome a substantial challenge to learn a single set of convolutional filters capable of distinguishing between land and ocean and effectively parameterizing the near-surface conditions leading to vastly different diurnal temperature regimes in summer and winter. The economy of the DLWP approach, which uses just four input variables and three prescribed fields, may be contrasted with the formulations in operational NWP and climate models where 2-m temperature is diagnosed using complex physical parameterizations for radiation, boundary-layer turbulence, and land- and ocean-surface interactions, all of which must be highly tuned for the model.

The T63 implementation of the IFS model, with a horizontal resolution similar to that of our cube-sphere grids, does clearly outperform our DLWP model.  It should be noted that the T63 IFS model improves on the T42 implementation not only through the use of higher horizontal resolution, but also by increasing the number of vertical levels to 137.  Unsurprisingly, the very high resolution operational S2S model also outperforms our DLWP model and the T63 IFS.  

Although our DLWP model lags the performance of a high-resolution operational NWP model by about 2--3 days of forecast lead time relative to climatology, it does have one significant advantage: computational speed. 
After a one-time computational cost of 2--3 days for training on a single NVidia Tesla V100 GPU, our DLWP model can produce a global four-week forecast in less than two tenths of a second.
At this speed one could generate a 1000-member ensemble of one-month forecasts in about three minutes. 
In contrast, the full dynamical IFS model at approximately  equivalent T63 horizontal resolution, run albeit somewhat inefficiently on a 36-core computing node, requires nearly 24 minutes to produce a single four-week forecast (Peter Dueben, personal communication), or about 16 days for the same 1000-member ensemble forecast. 
Operationally, ECMWF, despite vast supercomputing resources, only runs two-month-long S2S model forecasts twice weekly with 11 ensemble members. 
While DLWP models are likely to grow in complexity as they strive for better accuracy through the addition of more atmospheric variables, they hold great promise as a way of achieving the combination of speed and performance needed for very-large-ensemble weather forecasting.

%

\section{Acknowledgments}
The authors have greatly benefited from communications with Peter Dueben and Paul Ullrich. Peter Dueben provided the T42 and T63 IFS forecasts and guidance on the model parameters. Paul Ullrich provided valuable guidance on using the Tempest-Remap software. J. A. Weyn and D. R. Durran's contributions to this research were funded by Grant N00014-17-1-2660 from the Office of Naval Research (ONR). J. A. Weyn was also supported by a National Defense Science and Engineering Graduate (NDSEG) fellowship from the Department of Defense (DoD). Computational resources were provided by Microsoft Azure via a grant from Microsoft's AI for Earth program.


\bibliographystyle{abbrv}
\bibliography{ml}

\end{document}